\documentclass[apj]{emulateapj}

\usepackage{graphicx}
\usepackage{txfonts}
\usepackage[]{hyperref}

\bibpunct{(}{)}{;}{a}{}{,} 

\usepackage{natbib}
\usepackage{color}
\usepackage{multirow} 	
\usepackage{relsize}

\newcommand{\qm}[1]{``#1''}

\def\sss{\scriptscriptstyle}
\def\U{{\sss \!\mathrm{U}}}
\def\L{{\sss \!\mathrm{L}}}
\def\K{{\sss \!\mathrm{K}}}

\def\P{{\sss \!P}}
\def\S{{\sss \!S}}
\def\I{{\sss \!I}}
\def\C{{\sss \!C}}

\def\O{{\sss \!O}}
\def\R{{\sss \!R}}

\def\nur{\nu_\mathrm{r}}

\def\nuL{\nu_\L}
\def\nuU{\nu_\U}
\def\nuK{\nu_\K}

\def\nuP{\nu_\P}

\def\RISCO{\R\I\S\C\O}

\def\RISCO{r_{\mathrm{\RISCO}}}

\shorttitle{Mass and rapid variability of accreting compact objects}
\shortauthors{T\"{o}r\"{o}k et al.}

\begin{document}



\title{Simple analytic formula relating the mass and spin of accreting compact objects to their rapid X-ray variability}

\author{Gabriel~T\"{o}r\"{o}k, Andrea~Kotrlov\'{a}, Monika Matuszkov\'{a}, Kate\v{r}ina~Klimovi\v{c}ov\'{a},  Debora Lan\v{c}ov\'{a},\\ Gabriela Urbancov\'{a}, Eva~\v{S}r\'{a}mkov\'{a}
        }
        
        \affil{Research Centre for Computational Physics and Data Processing, {Institute of Physics}, Silesian University in Opava, Bezru\v{c}ovo n\'am.~13,\\CZ-746\,01 Opava, Czech Republic }
        
        \email{gabriel.torok@gmail.com}

\begin{abstract}
{
Following the previous  research on epicyclic oscillations of accretion disks around black holes (BHs) and neutron stars (NSs), a new  model of high-frequency quasi-periodic oscillations (QPOs) has been proposed (CT model), which deals with oscillations of fluid in marginally overflowing accretion tori (i.e., tori terminated by cusps). According to preliminary investigations, the model provides better fits of the NS QPO data compared to the relativistic precession (RP) model. It also implies a significantly higher upper limit on the Galactic microquasars BH spin. A short analytic formula has been noticed to well reproduce the model's predictions on the QPO frequencies in Schwarzschild spacetimes. Here we derive an extended version of this formula that applies to rotating compact objects. We start with the consideration of Kerr spacetimes and derive a formula that is not restricted to a particular specific angular momentum distribution of the inner accretion flow, such as Keplerian or constant. Finally, we consider Hartle-Thorne spacetimes and include corrections implied by the NS oblateness. For a particular choice of a single parameter, our relation  provides frequencies predicted by the CT model. For another value, it provides frequencies predicted by the RP model. We conclude that the formula is well applicable for rotating oblateness NSs and both models. We briefly illustrate application of our simple formula on several NS sources and confirm the expectation that the CT model is compatible with realistic values of the NS mass and provides better fits of data than the RP model. 
}

\end{abstract}


\section{Introduction}\label{Section:intro}

Accreting compact sources such as low-mass X-ray binaries (LMXBs) and active galactic nuclei (AGN) provide a unique opportunity to explore the effects associated with strong gravity in black hole (BH) and neutron star (NS) systems where they may also serve as a good tool for the exploration of supra-dense matter \citep[][]{kli:2006,Lew-etal:book:XrB}. There is a common aim within the large astrophysical community to relate the mass and spin of the compact objects to their spectral and timing behaviour. In this paper, we focus on the rapid X-ray variability and its models.

\begin{figure*}

\begin{center}
\noindent
\vspace{2ex}
a) \hfill  b) \hspace{1ex} \hfill ${\phantom{b}}$
\vspace{45ex}

\noindent
c) \hfill  d) \hspace{1ex} \hfill ${\phantom{b}}$
\vspace{45ex}

\noindent
e) \hfill  f) \hspace{1ex} \hfill ${\phantom{b}}$
\vspace{-99ex}

\noindent
\includegraphics[width=0.98\hsize]{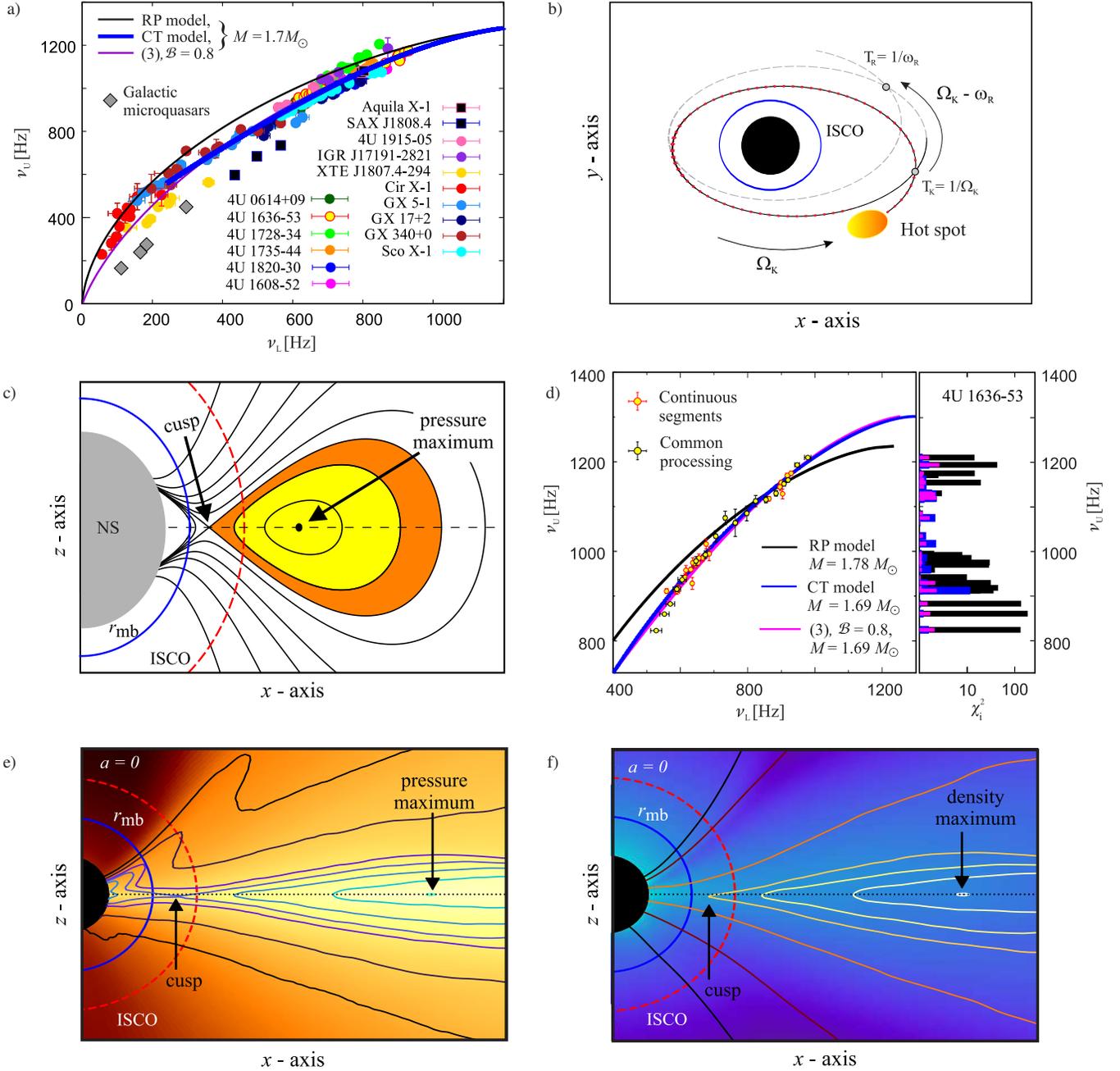}
\end{center}
\flushleft
\vspace{-5ex}

\noindent
\caption{a) The data of several sources and examples of  the expected frequency relations that are drawn for $M=1.7M_{\odot}$. The expected frequency relations are drawn for a non-rotating NS. b) Sketch of the trajectory of a test particle on a slightly eccentric orbit that plays a crucial role in the hot-spot interpretation of QPOs.  c) Topology of equipotential surfaces that determine the spatial distribution of fluid in thick disks. The orange (along with yellow) region corresponds to a torus with a cusp. The cusp is situated between the ISCO (the marginally stable circular orbit) and the marginally bound circular orbit (denoted as $r_{\mathrm{mb}}$). d) The data of the atoll source 4U~1636-53 \citep{bar-etal:2005,bar-etal:2005:b} and their best fits for non-rotating NS. For the sake of clarity, the data-set which corresponds to the individual continuous observations is compared to the data-set associated with the common processing of all observations. The right panel indicates individual contributions of each datapoint to the total chi-square value of a given fit \citep[see][for details]{tor-etal:2016:MNRAS}. e) The equipressure contours seen within general relativistic three-dimensional global radiative magnetohydrodynamic simulation of \citet{lan-etal:2019:ApJ} who have reported on a new class of realistic solutions of black hole accretion flows -- the so-called puffy accretion disks. f) Equidensity profiles corresponding to panel e.}
\label{figure:correlations}
\end{figure*}

The high-frequency part of the power density spectra (PDS) of many sources reveals more or less sharp peaks that are called the high-frequency quasi-periodic oscillations (HF QPOs). Commonly, the HF QPOs seem to have frequencies close to those of the orbital motion in the innermost part of a given accreting system. Detections of elusive HF QPO peaks in BH LMXB sources have been reported at rather constant frequencies, which tend to appear in ratios of small natural numbers \citep[often including a 3:2 ratio][]{abr-klu:2001,Rem-etal:2002:,mcc-rem:2006}. The observed HF QPOs are however very weak and the overall picture can be more complex \citep[][]{bel-etal:2012,bel-alt:2013,var-rod:2018}. In NS sources, HF QPOs are commonly referred to as twin-peak QPOs because they often appear in pairs observed simultaneously at the upper and lower QPO frequency, $\nuU>\nuL$. Notably, robust correlations are observed between the frequencies of twin-peak QPOs. Each source reveals its specific frequency correlation, $\nu_\U = \nu_\U(\nu_\L)$, although the sources roughly follow a common pattern \citep[][]{psa-etal:1999a,abr-etal:2005:AN,abr-etal:2005:RAG,zha-etal:2006}. This is illustrated in Figure~\ref{figure:correlations}a where we show the frequencies of QPOs in the 3:2 frequency ratio observed in Galactic microquasars along with the HF QPO correlations in a group of 14 NS sources.\footnote{For the sake of simplicity, we often use the shorter term "QPOs" instead of "HF QPOs" or "twin-peak QPOs" throughout the paper.} The data used in the figure come from the works of \citet{bar-etal:2005,bar-etal:2005:b,boi-etal:2000:,alt-etal:2010:,lin-etal:2005:,klis-etal:1997:,bou-etal:2006:,Hom-etal:2002:,Jon-etal:2000:,jon-etal:2002:} - NSs, and \citet{stroh:2001:,Rem-etal:2002:,hom-etal:2003,rem-etal:2003:} - BHs.

At present, it is commonly accepted that QPOs originate in the innermost parts of accreting systems, mostly likely being related to orbital motion. This view is supported by spectroscopic arguments and the similarity between the frequencies of QPOs and those of the orbital motion. There is, however, no commonly accepted QPO model. In fact, it has not even yet been resolved whether the generic mechanism is the same for both classes of the sources \citep[e.g.][]{kli:2006,men-bel:2021}.


The kinematics of the orbital motion allows for a straightforward consideration of  variability that is connected to inhomogeneites propagating in the accretion flow. Even before the era of the Rossi X-ray Timing Explorer QPO observations \citep{klis:1999:}, it was proposed that inhomogeneities moving from the disk to NS surface can produce imprints in the observed variability, which can be used to determine the NS properties \citep{klu-etal:1990:}. Later, several \qm{kinematic} models of QPOs were proposed, the most often quoted one being the \qm{relativistic precession} model (hereafter the RP model) introduced in a series of papers by \cite{ste-vie:1998a,ste-vie:1999,ste-vie:2002} and \cite{mor-ste:1999} \citep[see also][]{abr-etal:1992}. A similar concept is represented by models that include interactions between the disk and the surface or magnetosphere of rotating NSs \citep[][]{stroj-etal:1996:,mil-etal:1998a,psa-etal:1999c,hua-etal:2016:,wand-etal:2018:,wan-etal:2020:}. Another example of this type of models was discussed by \cite{cad-etal:2008}, \cite{kos-etal:2009}, and \cite{ger-etal:2009}. They introduced a scenario in which the QPOs are generated by a \qm{tidal disruption} (TD) of large accreted inhomogeneities.

Another, qualitatively rather different, possibility relies on the consideration of a collective motion of the accreted matter, in particular some fluid accretion disk oscillatory modes. Such concept has been explored for both thin (diskoseismology) and thick disks. Within the former case, a possible QPO mechanism has been suggested based on the early studies of \cite{kat-fuk:1980:}, \cite{now-etal:1991:}, \cite{kat-etal:1990:}, \cite{kat-hon:1991:}, \cite{now-etal:1992:}, and others. Reviews of the pioneering research can be found in \cite{wag:1999}, \cite{kat:2001}, while several subsequent papers were also published \citep[][]{wag-etal:2001,wag-etal:2012:,kat-mach:2020:,smi-etal:2021}. As for the latter, a large collection of papers has been published suggesting that the QPOs are related to oscillations of accretion tori \citep[][]{abr-klu:2001,rez-etal:2003,bur-etal:2004:APJ,bur:2005,abr-etal:2006,bla-etal:2006:, Ingram+Done:2010,fra-etal:2016,ave-etal:2017,mis-etal:2017:}. In the framework of fluid flow oscillations, based on the evidence for the appearance of ratios of small natural numbers, \cite{klu-abr:2001} have introduced the idea of a non-linear resonant coupling between different pairs of disk oscillation modes. This idea has been further extensively discussed \citep{abr-klu:2001,abr-etal:2003c,abr-etal:2003:d,klu-etal:2004,lee-etal:2004:,tor-etal:2005,abr-etal:2005:,pet:2005a,hor-etal:2006:,hor-etal:2009}. Among these often mentioned models, there are several alternative explanations, such as QPOs arising from shocks in advective accretion flows \citep{chak-tit-lev:1995:,chak-etal:1997,Le-etal:2016}, the Rossby waves instability \citep{li-fin-etal:2000:APJ:,Vin-etal:2013:}, or magnetic reconnection between the bounderies of the disk \citep{Zhao-etal:2009:,hua-etal:2013:}. While we do not intend to provide here a full review of the QPO models, we note that numerous modifications of models discussed in this paper exist along with other different concepts \citep[e.g.,][]{tit-ken:2002,rod-etal:2002:,zha:2004,wan-etal:2008,muk:2009,stu-kot:2009:,bach-etal:2010,don-etal:2011,stu-etal:2013,stu-kol:2014:,kol-etal:2015:,ger:2017:,frag:2020:,stu-etal:2020:,wan-etal:2020}. 

Having briefly illustrated the large collection of ideas proposed to explain the QPO phenomenon, in this work we focus solely on two particular QPO models. Following our previous study \citep[][]{tor-etal:2016:MNRAS,tor-etal:2016:ApJ,tor-etal:2018:MNRAS,tor-etal:2019}, we derive a simple analytic formula relating the QPO frequencies to parameters of rotating compact objects within the framework of these models. Finally, we apply the formula to data of several NS sources and compare the predictions of the models.

\section{QPO models under consideration}

We focus on two particular QPO models that deal with orbital motion of the accreted fluid. First is the above mentioned RP model, which in its usual form incorporates the assumption that the observed rapid X-ray variability originates in the local orbital motion of hot inhomogeneities orbiting in the innermost parts of the accretion disk, such as blobs or vortices \citep[see][]{klu-etal:1990:,abr-etal:1992,ste-vie:1998,ste-vie:1999}. 

The RP model has been used and quoted in numerous studies. It frequently serves as a rough tool for estimation of compact object mass based on its variability  \citep[e.g.,][and references therein]{bou-etal:2006:,bar-etal:2006,bou-etal:2007:,barr-etal:2008,bout-etal:2008:,tor-etal:2010:ApJ,lin-etal:2011,mot-bel-ste:2014,bui-etal:2019:MNRAS,mas-etal:2020:}. The relation between QPO frequencies postulated within the model has been shown to roughly match the NS sources data \citep[][see Figure~\ref{figure:correlations}a]{mor-ste:1999:, ste-vie:1998,ste-vie:1999,bel-etal:2007,lin-etal:2011,tor-etal:2012,tor-etal:2016:ApJ}. It is however highly questionable whether the local motion of hot spots can be responsible for the observed QPOs high amplitudes and coherence times \citep[][]{bar-etal:2005:MNRAS:357,bar-etal:2005}.

The other QPO model under consideration, which was proposed recently by \citet{tor-etal:2016:MNRAS}, assumes marginally overflowing accretion tori \citep[see the works of][for a broader context]{abr-klu:2001,klu-abr:2001,rez-etal:2003,bla-etal:2006:,ave-etal:2018}. This concept, to which we in next refer as the CT model, was suggested as a disk-oscillation-based alternative to the RP model. It utilizes the expectation that toroidal structures and cusps are likely to appear in real accretion flows, in which case the overall accretion rate through the inner edge of the disk could be strongly modulated by the torus oscillations \citep{koz-jar-abr:1978,Abr-Jar-Sik:1978:ASTRA:,pac-abr:1982,hor:2005:AN:,abr-etal:2007,par-etal:2017:}. A sketch of the marginally overflowing acretion torus geometry is shown in Figure~\ref{figure:correlations}c. 

The concept of the CT model stems from the outputs of previous studies devoted to oscillations and stability properties of fluid tori. These were initiated by \citet{Pap-Pri:1984:MONNR:}, who investigated the global linear stability of fluid tori with respect to non-axisymmetric perturbations and, considering small linear perturbations to the torus equilibrium, derived a single partial differential equation governing the linear dynamics of oscillations of a Newtonian constant specific angular momentum torus. Later, a general relativistic form of the Papaloizou--Pringle (PP) equation was introduced by \citet{abr-etal:2006} and \citet{bla-etal:2006:}. Since this equation cannot be fully solved analytically,  \citet{str-sram:2009} and {\citet{fra-etal:2016}} used a perturbation method to derive fully general relativistic formulae determining frequencies of the axisymmetric and non-axisymmetric radial and vertical epicyclic modes in a slightly non-slender constant specific angular momentum torus within a second-order accuracy in the torus thickness. Their outputs were subsequently applied in \citet{tor-etal:2016:MNRAS}.

The CT model provides generally better fits of the NS data than the RP model. This is depicted in Figure~\ref{figure:correlations}d. This finding is rather independent on the NS spin \citep[][]{tor-etal:2016:MNRAS,tor-etal:2016:ApJ}. The model also likely predicts a lower NS mass compared to the RP model, which, in some cases, implies a mass estimate that is too high \citep{tor-etal:2016:MNRAS,tor-etal:2018:MNRAS,tor-etal:2019}. Moreover, the upper limit on the Galactic microquasars spin given by this model is significantly higher than in the RP model's case, namely $j\sim0.75$ vs. $j\sim 0.55$. This is in better agreement with the spectral spin estimates \citep[][]{kot-etal:2020:AA:}.

An overview of the physical assumptions of the model, as well as all appropriated references, can be found in the studies of \citet{tor-etal:2016:MNRAS} and \cite{kot-etal:2020:AA:}. Similarly to the RP model, the currently applied concept of inner torus displaying a cusp is very simplified compared to real accretion flows. Note that, at their present stage, both considered models do not explain why the assumed modes of the accreted matter motion are well observed while other, similar, modes are not. Nevertheless, the CT model assumes a global motion of disk fluid as oppose to local test particle motion considered in the RP model. Moreover, the structure of the inner accretion flow observed in general relativistic radiation magnetohydrodynamic (GRRMHD) simulations often resembles that assumed within the model. This is illustrated in Figures~\ref{figure:correlations}e and f. Although there still is a rather long way to go in the development of the GRMHD simulations in order to become able to study the possible QPO mechanism \citep[see, however,][]{mus-etal:2022:}, the similarity between the structures supports the expectation that the adopted approximation of the accretion flow posed by marginally overflowing tori allows for determination of oscillatory frequencies close to those of accretion flows in real systems.

\begin{figure*}
\begin{center}
\hspace{2ex} a) \hfill  ~~b)  \hfill ~~c) \hfill ~~d) \hfill ${\phantom{e}}$
\medskip

\noindent
\includegraphics[width=\hsize]{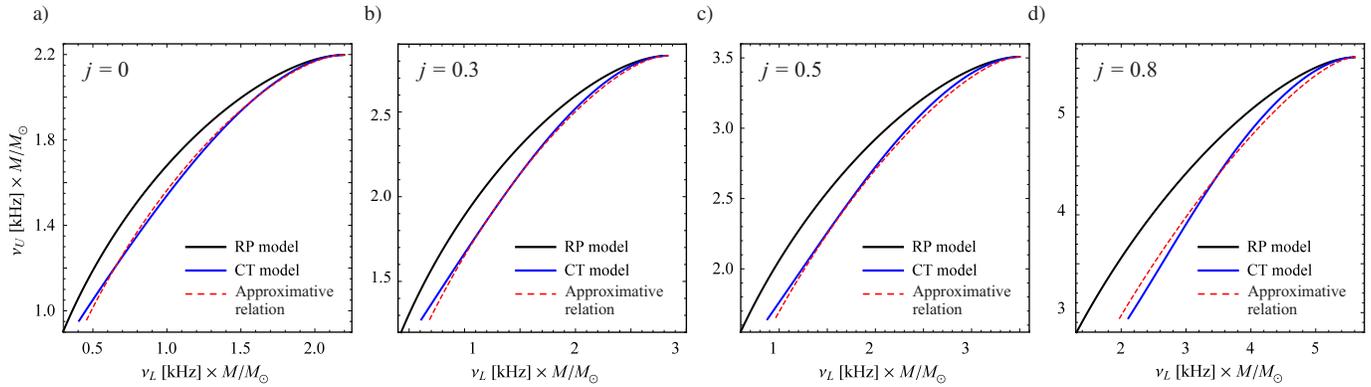}
\end{center}
\flushleft
\vspace{-3ex}

\noindent
\caption{Comparison between expected QPO frequencies (Kerr spacetimes). The individual curves denote frequencies given by analytical prescription for RP model, numerically calculated for CT model, and those given by relation~(\ref{equation:RP:Kerr}) for $\mathrm{k}=-0.2$.}
\label{figure:relations}
\end{figure*}

\section{Frequencies of QPOs within the RP model}

Within the RP model, the frequencies of the two observed QPOs are given by the Keplerian frequency $\nuK$ and the relativistic precession frequency $\nuP$ of a slightly perturbed circular geodesic motion occuring at an arbitrary QPO excitation orbital radius $r_{0}$ (see Figure~\ref{figure:correlations}b),
\begin{equation}
\label{equation:RP}
\nuU(r_{0}) = \nuK(r_{0}), \quad \nuL(r_{0}) = \nuP(r_{0})\,.
\end{equation}
The precession frequency equals to a difference between the Keplerian and the radial epicyclic frequency,
\begin{equation}
\nuP(r_{0}) = \nuK(r_{0}) - \nur(r_{0})\,.
\end{equation}

For a non-rotating relativistic compact star with the external spacetime given by  the Schwarzschild geometry, the relation between the QPO frequencies implied by the RP model, i.e. by Equation (\ref{equation:RP}), can be written as \citep{ste-vie:1999}
\begin{equation}
\label{equation:the-one}  
\nuL= \nuU\left(1 - {\mathcal{B}}\sqrt{1 - \left(\nuU/\nu_0\right)^{2/3}}\right),
\end{equation}
where $\mathcal{B}=1$ and $\nu_0$ equal to the Keplerian frequency at the innermost stable circular orbit (ISCO). The ISCO frequency is given solely by the gravitational mass $M$,
\begin{equation}
\nu_{\mathrm{0}}=\mathcal{F}\frac{1}{6^{3/2}}, \quad \mathcal{F} \equiv \mathrm{c}^3/(2\pi \mathrm{G}M)\,.
\end{equation}

\section{Frequencies of QPOs within the CT model}

For the CT model, the relation between the QPO frequencies in the Schwarzschild spacetimes also depends purely on $M$ and can be written in the following implicit form \citep{tor-etal:2016:MNRAS}:
\begin{equation}
\label{equation:CT}
\nuU(r_{0}) = \nuK(r_{0}), \quad \nuL(r_{0}) = \nu_{\mathrm{r},\,m=-1}\,,
\end{equation}
where $r_{0}$ denotes the torus centre where the density of fluid peaks, $\nuK$ determines the torus centre corotation frequency, and $\nu_{\mathrm{r},\,\mathrm{m}=-1}$ equals to the frequency of the first non-axisymmetric radial epicyclic mode calculated for the marginally overflowing torus (i.e., the torus that forms a cusp). Contrary to the RP model, which considers a geodesic test particle precession, this frequency corresponds to precession of the whole fluid flow, which can strongly modulate accretion rate through the boundary layer. We note that, in contrast to the $m=-1$ mode, the modulation mechanism behind the imprints of the Keplerian frequency has not yet been fully resolved within the model's framework. We expect that unstable corotation oscillatory modes along with the dynamics of inhomogeneities formed in the flow may contribute to the accretion rate modulation. The  timescale of disk oscillations is more than five orders of magnitude shorter than the typical integration time required to well identify the two peaks in the PDS. Consequently, an overall emphasis of frequencies that are close or equal to the Keplerian frequency at the torus centre can arise in the X-ray PDS, at least when small tori close to the ISCO are considered. Within the model framework, small tori likely correspond to the case of high QPO frequencies observed in the atoll sources. Our expectation is also in agreement with the results of numerical simulations of the (PP) instability. It has been shown that non-linear evolution of the instability with azimuthal wavenumber $m$ leads to fragmentation of the initial torus configuration into a configuration with $m$ nearly disconnected "planets" \citep[see the early studies of][]{gold-etal:1986:,goo-etal:1987:,nar-etal:1987:}. Since the instabilities can be supressed by accretion through the cusp \citep[][]{bla:1987}, the upper and lower QPO can alternate each other. Nevertheless, more investigation is needed to obtain a comprehensive understanding of the physical mechanism behind the upper QPO.


There is no explicit analytical evaluation of $\nuU(\nuL)$ function and the relation (\ref{equation:CT}) must be solved numerically. It has been however noticed by \citet{tor-etal:2018:MNRAS} that there is a solid analytic approximation -- the numerical solution nearly coincides with relation (\ref{equation:the-one}) when $\mathcal{B}=0.8$. This is illustrated in Figure~\ref{figure:correlations}b.

\section{Rotating compact stars}
\label{Section:Kerr}

It has been noticed by \citet{tor-etal:2010:ApJ} that in Kerr spacetimes characterized by the $j\equiv\mathrm{c}J/(\mathrm{G}M^2)$ rotational parameter, the relation between the QPO frequencies implied by the RP model can be expressed as
\begin{eqnarray}\label{equation:RP:Kerr}
\nuL = \nuU\left\{1 - {\mathcal{B}}\left[1 + \frac{8j\nuU}{\mathcal{F} - j\nuU} - 6\left(\frac{\nuU}{\mathcal{F} - j\nu_U}\right)^{2/3}
\right.\right. \nonumber \\ 
\left.\left.-  
3j^2\left(\frac{\nuU}{\mathcal{F} - j\nuU}\right)^{4/3} \right]^{1/2}\right\}\,,
\end{eqnarray}
when one sets $\mathcal{B}=1$.

Based on the analogy to non-rotating stars, relation (\ref{equation:RP:Kerr}) for a particular choice of $\mathcal{B}(j)$ can be expected to reproduce the numerically calculated frequency relation given by the CT model. Having this intention in mind, we presume a simple linear prescription,
\begin{equation}
\label{equation:linear}
\mathcal{B}(j) = \mathrm{k}j+0.8\,
\end{equation}
which results in $\mathcal{B}=0.8$ for the $j=0$ limit.

We made a comparison between the predictions of relation (\ref{equation:RP:Kerr}) and the CT model predictions. Following the approach of \cite{tor-etal:2016:MNRAS}, we use the results of analytical calculations of the perturbative solution of the PP equation and compute the numerical solution of the dependence between the expected QPO frequencies. As noticed by \cite{hor-etal:2017:}, the applied analytic perturbative solution of the PP equation providing the frequency of the radial epicyclic modes is in a good agreement with the appropriate numerical solution. The whole set of formulae necessary for the numerical calculations is given by fairly long expressions, we therefore provide their explicit
 form in a Wolfram Mathematica notebook \citep[see][]{kot-etal:2020:AA:}. 
A good match between the analytical prescription and the numerically calculated predictions is found for $\mathrm{k} = -0.2$ and illustrated in Figure~\ref{figure:relations}.

We note that the applicability of our result to rapidly rotating BHs is limited. The so far performed numerical calculations of frequencies given by the CT model utilize a perturbative approach valid within the second-order accuracy in torus thickness. Within this approach, the calculations are for high spins very sensitive to small changes in the torus thickness and the rotational parameter. Full numerical investigation of the PP equation, which determines the epicyclic mode frequencies, will be needed for rapidly rotating BHs.

\begin{figure*}
\begin{center}
a) \hfill  ~~~~b) \hfill ~~~~c) \hfill ${\phantom{b}}$\\
\includegraphics[width=1\hsize]{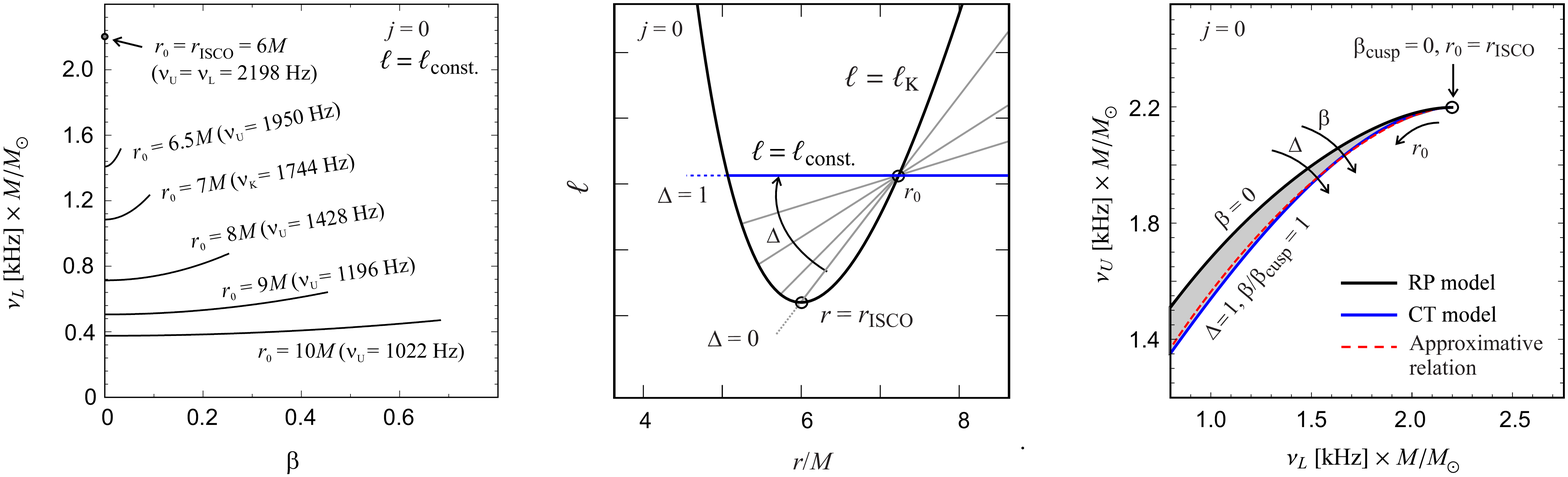}
\end{center}

\noindent
\caption{a) Behaviour of the $\mathrm{m}=-1$ radial oscillation mode frequency $\nu_{\mathrm{r},\,\mathrm{m}=-1}$. Frequency values quoted in the labels are evaluated for one solar mass star. Each curve starts at the frequency corresponding to a configuration of a slender torus and terminates at the frequency corresponding to a configuration of a torus with the largest possible radial extent (torus that has a cusp). Note that tori with $r_0$ close to $r_0=r_{\mathrm{ISCO}}$ can have a much smaller radial extent than those located at high $r_0$, since cusp tori terminate between the ISCO and the marginally bound orbit. b) Sketch of various angular momentum distributions of the accreted fluid represented by Keplerian distribution and a family of linear distributions characterized by a $\Delta$-parameter. c) The expected QPO frequencies. Shaded area denotes frequency range corresponding to all combinations of $\Delta$-parameter, $\Delta\in[0,\,1]$, and torus thickness $\beta$, $\beta\in[0,\,\beta_{\mathrm{cusp}}]$. \label{figure:distribution}}

\end{figure*}

\section{QPO frequencies and accretion flow angular momentum distribution}

Following the previous studies of \citet{str-sram:2009}, \citet{tor-etal:2016:MNRAS}, \citet{fra-etal:2016} and \citet{kot-etal:2020:AA:}, our simple formula relating the QPO frequencies and the compact objects mass and spin was derived under the specific consideration of tori with constant distribution of angular momentum $\ell$ of the accreted fluid. Nevertheless, it can be easily shown that the formula is of a more general importance since its validity is not limited to this particular $\ell$-prescription.

\subsection{Range of QPO frequencies}

If we replace the $\mathcal{B}$ factor in equation (\ref{equation:RP:Kerr}) by unity, we obtain exactly the prediction based on the test particle motion. This case also describes the scenario in which the angular momentum distribution of the accreted fluid is Keplerian. It is a limit case, in which the cross-section of the oscillating torus is infinitely small (a parameter determining the torus thickness, $\beta$, goes to zero).\footnote{The beta parameter describes the torus thickness and its relative radial extent such that, at a given $r_0$, a higher beta corresponds to a higher torus extent. It is given as $
    \beta =
{\sqrt{2n}\,c_\mathrm{s}}/{{r_\mathrm{0}}\Omega_\mathrm{0} u^t_\mathrm{0}},
$
where $n$ is the polytropic index, $c_\mathrm{s}$ the polytropic sound speed, $u^t_\mathrm{0}$ denotes the contravariant time component of the four-velocity, and $\Omega_\mathrm{0}$ indicates the angular velocity of the flow. All these quantities are defined at the centre of the torus, $r=r_\mathrm{0}$.}

When the $\mathcal{B}$ factor is taken into account, we come up with a scenario in which the torus has its maximal possible size ($\beta=\beta_{\mathrm{cusp}}$). As shown by \citet{str-sram:2009}, the $\mathrm{m}=-1$ radial mode frequency evolves as a monotonic function of the torus thickness (see Figure~\ref{figure:distribution}a for illustration). Accordingly, the black curve (marked as the RP model) and the coloured curves (CT model) in Figure~\ref{figure:relations} describe the two extreme predictions of the QPO frequencies given by equation (\ref{equation:RP:Kerr}). The area between these curves covers the whole range of the QPO frequencies determined under the consideration of constant $\ell$ and any $\beta$. When we put
\begin{equation}
\mathcal{B}= 1-0.2(1 + j)\frac{\beta}{\beta_{\mathrm{cusp}}},
\end{equation}
we obtain a continuous set of curves that cover the area between the limiting curves given by $\beta = 0$ and $\beta = \beta_{\mathrm{cusp}}$.

\subsection{Generalization for non-constant angular momentum distributions}

The above consideration can be extended to a more general picture. In panel a) of Figure~\ref{figure:distribution}, we show the behaviour of the $\mathrm{m}=-1$ radial frequency (the expected lower QPO frequency) for the particular case of $\ell=\ell_{\mathrm{const}}$. One should note the increasing monotonic behaviour of the curves. When $\beta$ increases, the $m=0$ mode frequency decreases and the $m=-1$ mode frequency increases getting closer to the Keplerian frequency. The trend of the $m=-1$ mode frequency rising with growing size of the oscillating structure persists for less simplified situations as well.
In panels b) and c) of Figure~\ref{figure:distribution}, we show a sketch of a possible parametrization of linear angular momentum distributions and its projection in the plane of the expected QPO frequencies. Clearly, taking into account the presumption of a monotonic behaviour of the $\nu_{\mathrm{r},\,\mathrm{m}=-1}(\beta)$ function, any linear prescription for the angular momentum distribution, and all possible torus thicknesses from the $\beta\in[0,\,\beta_{\mathrm{cusp}}]$ range, the expected QPO frequencies should fall into the range denoted by the shaded area. An analogical consideration also applies to non-linear distributions.  

Overal, the narrow range between the two extremal curves given by equation (\ref{equation:finalI}), which is indicated by the shaded area in Figure~\ref{figure:distribution}b, represents a rather general limit on the QPO frequencies valid for a variety of plausible angular momentum distributions.

\section{NSs and their oblatness}

Formula (\ref{equation:RP:Kerr}) is valid for Kerr spacetimes relevant for black hole sources. Within a reasonable accuracy, it can be applied to neutron star sources as well provided that the NS mass is high. Considering the restrictions on NS quadrupole moment $q\equiv Q/ M_{0}^3 $ given by present NS equations of state and consequent implications on the orbital frequencies \citep{Urb-etal:2013:,urb-etal:2019}, we can estimate the uncertainty in our formula valid for most of the available NS data. For high NS masses (typically $M\gtrsim 2M_{\odot}$) and spins corresponding up to $j\sim0.3$, the uncertainty in NS mass induced within our formula by the quadrupole moment should not be higher than $3\%$. On the other hand, for low NS masses ($M\lesssim 1.4M_{\odot}$), this uncertainty may exceed the value of $10\%$. 

A modification of the formula that would be sufficiently valid for such less compact NS sources can be obtained assuming the Hartle-Thorne geometry \citep[][]{har:1967:,har-tho:1968:}, which applies to slowly rotating  neutron stars. One may expect that the impact of the quadrupole moment consideration on the relation between the QPO frequencies can be roughly included substituting the (rotational) $3j^2$ term in formula (\ref{equation:RP:Kerr}) by a simple dependency on $q$, $\mathcal{Q}=\mathcal{Q}(j,~q)$. In the limit of $q=j^2$, the Hartle-Thorne formula should coincide with those expressed in the Kerr spacetime \citep[e.g.,][]{urb-etal:2019} and there is
\begin{equation}
\mathcal{Q}(j,~q)=\mathcal{Q}(j,~j^2)=3j^2\,.
\end{equation}

In analogy to Section~\ref{Section:Kerr}, we use the results of analytical calculations of the perturbative solution of
PP equation and compute the numerical solution of the dependence between the expected QPO frequencies. Then we attempt to ﬁnd the best evaluation of the Q-term performing numerical calculations of the CT model frequencies in Hartle-Thorne spacetimes. We utilize the results of \cite{fra-etal:2016} and \cite{kot-etal:2020:AA:} and their extension to Hartle-Thorne spacetimes.

\begin{figure*}[ht]
\begin{center}

\noindent
\includegraphics[width=\hsize]{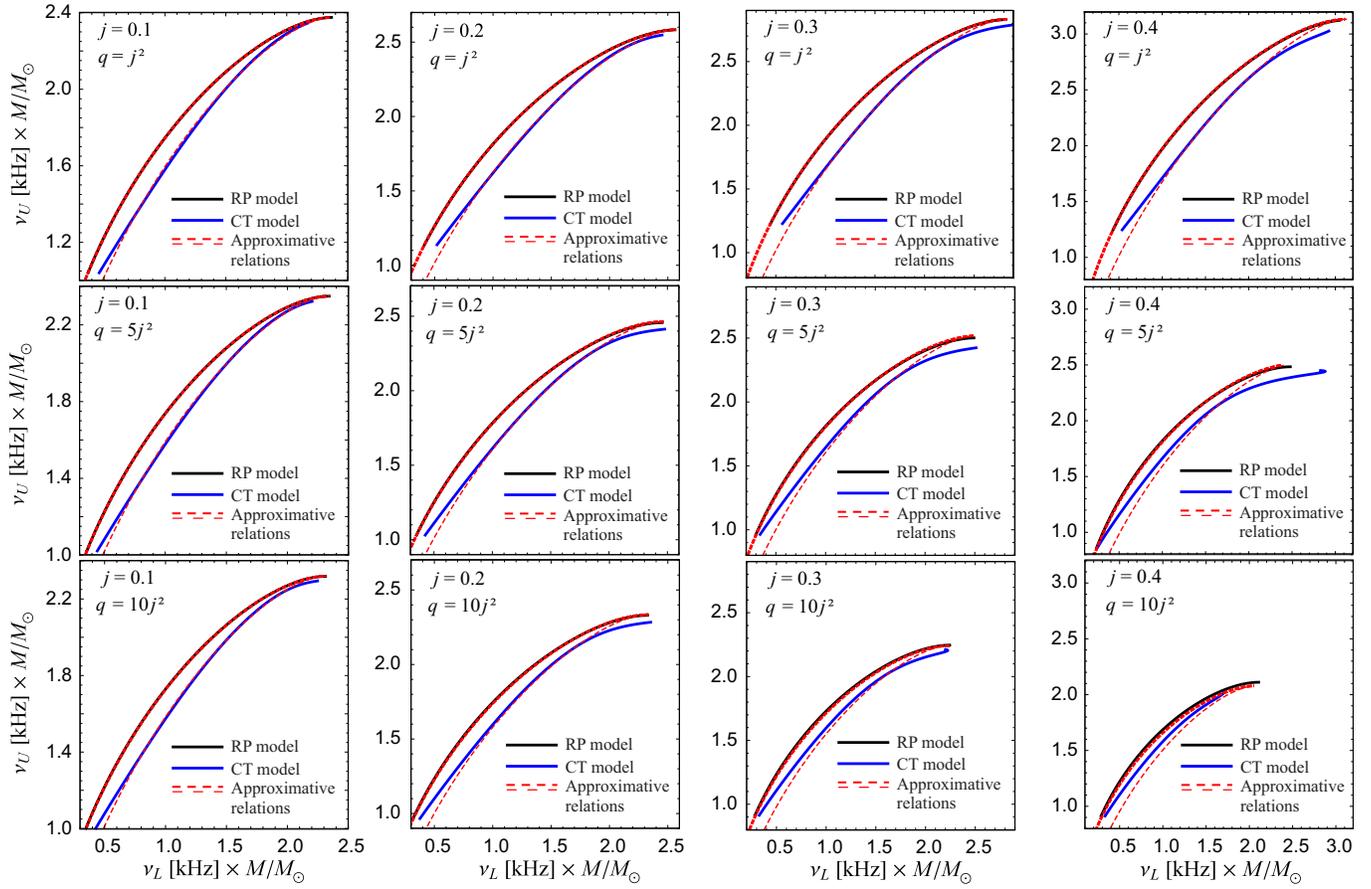}
\end{center}
\flushleft
\vspace{-0.5cm}

\noindent
\caption{Comparison between the expected QPO frequencies (Hartle-Thorne spacetimes). The continuous curves denote frequencies calculated numerically for the CT and RP models. The dashed curves denote frequencies calculated using the approximative relation for the CT ($\mathcal{B}=0.8-0.2j$) and RP ($\mathcal{B}=1$) models.}
\label{figure:HT}
\end{figure*}

We find that the particular term
\begin{equation}
\mathcal{Q}=\frac{1}{3}(17 q - 8 j^2 )\,
\end{equation}
well matches the numerical calculations. This is illustrated in Figure~\ref{figure:HT}. It is also clear from the Figure that for $\mathcal{B}=1$ we obtain the frequencies predicted by the RP model. We note that for high QPO frequencies, corresponding to radii close to ISCO, when $\nuL$ approaches $\nuU$, there are discrepancies between the examined relations. These follow from the limitations of the Hartle-Thorne approach, which is accurate up to the second-order terms in $q$. The inaccuracies however grow only when the difference between the two QPO frequencies, $\Delta\nu=(\nuU-\nuL)/\nuL$, is smaller than $10\%$.

\begin{figure*}[ht]
\begin{center}

\noindent
a) \hfill  b) \hspace{-5ex} \hfill ${\phantom{b}}$
\vspace{36ex}

\noindent
c) \hfill  d) \hspace{-5ex} \hfill ${\phantom{b}}$

\vspace{-41ex}

\noindent
\includegraphics[width=\hsize]{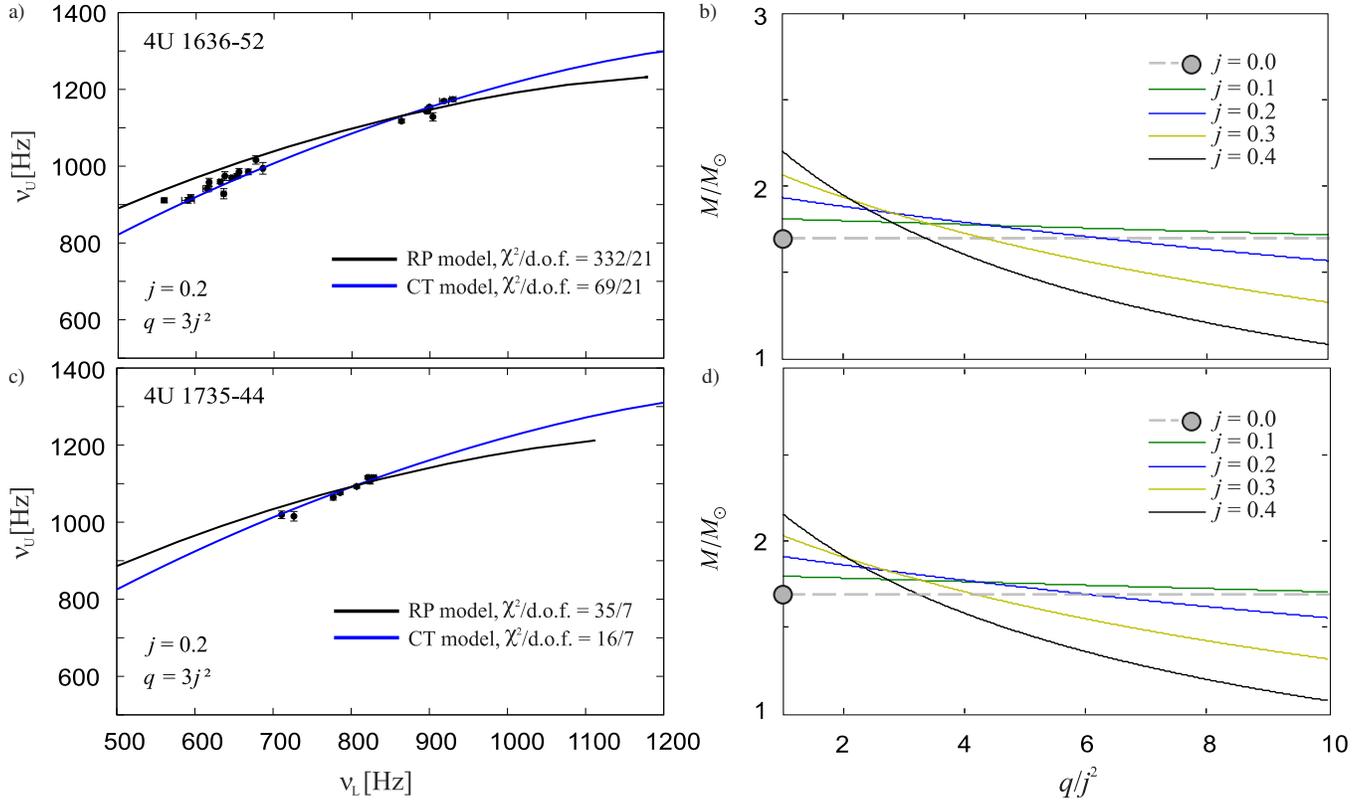}
\end{center}
\flushleft

\noindent
\caption{a) Best fits of the data of the 4U~1636-53 atoll source found for the RP and CT models and a particular choice of the NS spin and oblateness. For the other choices within the considered range of parameters, $j\in[0,\,0.4]$ and $q/j^2\in[1,\,10]$, the resulting fits are similar.
b) The best fitting mass corresponding to the CT model as it depends on $q/j^2$. c) The same as in panel a) but for the 4U~1735-44 atoll source. c) The same as in panel b) but for the 4U~1735-44 atoll source.}
\label{figure:HT2}
\end{figure*}

\section{Discussion and conclusions}\label{Section:conclusions}

For practical purposes, taking into account the ISCO frequency term for non-rotating stars, $\nuL=\nuU=\nu_0$, equation (\ref{equation:RP:Kerr}) can be further rewritten into a final compact form,
\begin{equation}
\label{equation:finalI}
\nu_{\mathrm{L}} = \nu_{\mathrm{U}}\left[1 - \mathcal{B}\sqrt{1 + 8j\mathcal{V}_0
- 6\mathcal{V}_0^{2/3}
- \mathcal{Q}\mathcal{V}_0^{4/3}}\right]\,,
\nonumber
\end{equation}
where
\begin{eqnarray}
\label{equation:finalII}
&&\mathcal{B}= 0.8 - 0.2j\,, \quad
\mathcal{V}_0 \equiv \frac{\nu_{\mathrm{U}}/\nu_{\mathrm{0}}}
{6^{3/2} -j\nu_{\mathrm{U}}/\nu_{\mathrm{0}}} \nonumber
\\
&& \nu_{\mathrm{0}}=2198\frac{\mathrm{M}_\odot}{M}\,,
\mathcal{Q}=\frac{1}{3}(17 q - 8 j^2)\,. \nonumber
\end{eqnarray}
For the above choice of $\mathcal{B}$, our relation with high accuracy provides frequencies predicted by the CT model. Choosing a constant $\mathcal{B}$, $\mathcal{B}=1$, it (almost exactly) provides frequencies predicted by the RP model. We therefore conclude it is applicable for both models in the case of rotating oblateness NSs. For $\mathcal{Q}=3j^2$, the relation reduces to the case of Kerr spacetimes describing rotating BHs.

\subsection{Application to the atoll source 4U~1636-53 and other NSs}

Following \cite{tor-etal:2016:MNRAS}, we apply relation \ref{equation:finalI} to the data of the atoll source 4U~1636-53. The main outputs of our investigation are illustrated in Figure~\ref{figure:HT2}. Figure~\ref{figure:HT2}a includes examples of the best fits given by the CT model. Fits given by the RP model are shown as well for the sake of comparison. Figure~\ref{figure:HT2}b depicts how the best fitting $M$ depends on $j$ and $q/j^2$. It shows that for very compact with $q/j^2\sim1$ the best fitting $M$ increases with increasing $j$, reaching values of $M\in[2,\,2.2]M_{\sun}$ for $j\in[0.2,\,0.4]M_{\sun}$. This is in agreement with the investigation of \cite{tor-etal:2016:MNRAS} limited to the case of Kerr spacetimes. On the other hand, for stars of high oblateness, $q/j^2>4$, the best fitting $M$ decreases with increasing $j$. For stars of moderate oblateness, $q/j^2\sim3$, there is only a very weak dependency on $j$ and the estimated mass is around $M=1.75M_{\sun}$.

The same investigation was performed for the atoll source 4U~1735-44. The results are illustrated in Figure~\ref{figure:HT2} showing a picture very similar to the 4U~1636-53 case. In analogy to the 4U~1636-53 case, we obtain fits better than those of the RP model and similar quadrupole moment dependence. For very compact stars, $q/j^2\sim1$, the best fitting $M$ increases with increasing $j$, reaching values of $M\in[1.9,\,2.2]M_{\sun}$ for $j\in[0.2,\,0.4]M_{\sun}$, while for stars of high oblateness, $q/j^2>4$, the best fitting $M$ decreases with increasing $j$. For stars of moderate oblateness, $q/j^2\sim3$, there is only a very weak dependency on $j$ and the estimated mass is around $M=1.9M_{\sun}$. In the same way, we investigated another four atoll sources with high amount of available data \citep[][]{bar-etal:2005,bar-etal:2005:b,tor-etal:2012}. Overall, we find that for stars of moderate oblateness, $q/j^2\sim3$, the mass should be within the interval of $M\in[1.6,\,1.9]M_{\sun}$.

These findings further confirm the expectation that the CT model not only fits the data better than the RP model, but is also compatible with realistic values of the NS mass.

\subsection{Caveats}

Our finding on the NS mass needs to be expanded to a larger set of sources, namely to a full confrontation of the parameters implied by the model and particular NS equations of state. It should be sufficient if this confrontation is carried out within the framework of the Hartle-Thorne spacetime for most sources and data except for very rapidly rotating sources and data with $\nuU/\nuL<1.2$. It is questionable whether the present relation can be applied to sources with very strong magnetic fields such as X-ray pulsars. The applicability of our result to rapidly rotating BHs has yet to be explored as well using the full numerical solution of the PP equation.

Despite these caveats, we conclude that the simple relation (\ref{equation:finalI}) can be useful for a brief estimation of mass and spin of accreting BHs and NSs.

\section*{Acknowledgments}
We thank Ji\v{r}\'{\i} Hor\'{a}k for very useful discussions. We also thank the anonymous referee for several suggestions that greatly helped to improve the paper. We acknowledge the Czech Science Foundation (GA\v{C}R) grant No.~21-06825X. We wish to thank internal grants of the Silesian University in Opava, SGS/12,13/2019. DL thanks the Student Grant Foundation of the Silesian University in Opava, Grant No. $\mathrm{SGF/1/2020}$, which has been carried out within the EU OPSRE project entitled ``Improving the quality of the internal grant scheme of the Silesian University in Opava'', reg. number: $\mathrm{CZ.02.2.69/0.0/0.0/19\_073/0016951}$. KK was also supported by the Czech grant LTC18058 and the COST Action PHAROS (CA 16124). We also acknowledge the ESF project reg. number $\mathrm{CZ.02.2.69/0.0/0.0/18\_054/0014696}$.

\end{document}